\RequirePackage{pdf14}

\documentclass[letterpaper,10pt]{article}
\pdfoutput=1
\usepackage[utf8]{inputenc}
\usepackage[usenames,dvipsnames,table]{xcolor}
\usepackage{graphicx}
\usepackage{caption}
\usepackage{subcaption}
\usepackage{array,setspace,mathrsfs,amsthm,amsfonts,amsmath,colonequals,mathtools,bbm}
\usepackage{./auxi/jheppub}
\usepackage{afterpage}
\usepackage{xspace}
\usepackage[normalem]{ulem}
\usepackage{cancel}
\usepackage{cleveref}
\usepackage{hyperref}
\usepackage{shuffle}

\newcommand{\beqa}{\begin{eqnarray}}
\newcommand{\eeqa}{\end{eqnarray}}
\newcommand{\beq}{\begin{equation}}
\newcommand{\eeq}{\end{equation}}

\newcommand{\SB}{\text{SB}} 
 \newcommand{\calV}{\mathcal{V}} \newcommand{\calH}{\mathcal{H}}
  \newcommand{\calN}{\mathcal{N}}
  \newcommand{\calP}{\mathcal{P}}
\newcommand{\calQ}{\mathcal{Q}} 
\newcommand{\mker}{\text{Ker}}\newcommand{\mim}{\text{Im}}

\makeatletter
\def\maketag@@@#1{\hbox{\m@th\normalfont\normalsize#1}}
\makeatother

\title{SymBuild: a package for the computation of  integrable symbols in
scattering amplitudes }

\author[1]{Vladimir Mitev,}
\author[1,2]{Yang Zhang}

\affiliation[1]{PRISMA Cluster of Excellence, Institut f\"ur Physik,
JGU Mainz,
Staudingerweg 7, 55128 Mainz, Germany}
\affiliation[2]{Institute for Theoretical Physics, ETH Z\"urich, CH 8093 Z\"urich, Switzerland}

\emailAdd{vmitev@uni-mainz.de}
\emailAdd{zhang@uni-mainz.de}

\preprint{MITP/18-079}

\bigskip
\abstract{
The article presents and documents the \textsf{Mathematica} package \textsf{SymBuild}. This package
implements the computation and manipulation of integrable symbols appearing in various calculations 
in high-energy scattering amplitudes. By using Gr\"obner bases, implementing various simplifications and by the potential utilization of the \text{C++} program \textsf{SpaSM}, integrable symbols in a large class of alphabets (including roots) can be computed to high weights.
}

\begin{document}
\setcounter{tocdepth}{2}
\maketitle
\setcounter{page}{1}





\section{Introduction}

An ancient idea in the field of scattering amplitudes, going back to the analytic S-matrix program \cite{Eden:1966dnq}, has been to determine, or bootstrap, a given scattering amplitude by using its physical and analytical properties. In some 1-loop cases, the constraints of perturbative unitarity and the known collinear/soft limits have been used to determine explicitly entires classes of $n$-particle scattering amplitudes \cite{Bern:1994zx}. A big impediment to the implementation of this method at higher loop orders has been the fact that higher-loop Feynman integrals 
are significantly less understood. However, once it was realized that a large class of Feynman integrals are given by iterated integrals which in many situations can be efficiently described via their\footnote{ These symbol techniques have been used in the mathematical community much earlier, see for example \cite{Zagier1991, Goncharov_1995,Lewin1991,  Gangl_2003}.} \textbf{symbols} \cite{Goncharov:2010jf}, further progress became possible. The idea behind the symbol is to encode the way in which the functions are defined from elementary integrands (called \textbf{letters}, the set of which is called the \textbf{alphabet}). Then, the symbol of an $L$-fold iterated integral (referred to as a \textbf{weight} $L$ function) encodes its analytic structure  such as branch surfaces and poles. 

As soon as one knows the symbol alphabet of a given scattering amplitude, one can write a very constraining ansatz for it. In many cases, the ansatz for a loop order $\ell$ amplitude is made out of a set of prefactors (that can be determined by computing leading singularities) that multiply a transcendental function of weight $2\ell$. Instead of working with the functions, one works with their symbols and reconstructs the full answer at the end. The set of symbols of all such transcendental functions can be constructed iteratively from the alphabet by solving a set of \textbf{integrability conditions}, which basically impose that that symbol represents a well-defined function, and the so-called $n$-entry conditions\footnote{Steinmann relations impose on the symbols a type of $n$-entry condition.}, which come from physical constraints on the branch surfaces. It is however a non-trivial linear algebra problem to compute the set of integrable symbols subject to the $n$-entry conditions, a task that the package introduced in this article should simplify. 

Once the set of integrable symbols is known, all one in principle needs to do is to solve a set of linear equations coming from collinear/soft constraints and others and hope that this proves enough to fix all unknown coefficients in the ansatz. 
This modern incarnation of the bootstrap program has been applied successfully to many cases, such as the 6- and 7-point amplitudes in planar $\calN=4$ super Yang-Mills (SYM) \cite{Dixon:2011nj, Dixon:2011pw, Dixon:2013eka, Dixon:2014iba, Dixon:2015iva, Caron-Huot:2016owq, Drummond:2014ffa, Dixon:2016nkn}, the computation of Feynman integrals \cite{Chicherin:2017dob, Henn:2018cdp,Chicherin:2018wes}, the QCD soft anomalous dimension \cite{Almelid:2017qju} or Higgs boson amplitudes \cite{Duhr:2012fh}. As the loop order $\ell$, and hence also the weight, increases, solving the integrability conditions becomes increasingly harder, since the size of the underlying linear algebra problem grows roughly as $M^{2\ell}$ where $M$ is the number of letters in the alphabet. It eventually becomes necessary to use tools that go beyond what a standard computer algebra program like \textsf{Mathematica} can do in order to compute the integrable symbols and then manipulate them. As an aside, not only is the computation of the integrable symbols for a given alphabet difficult, but even the comparatively easier task of computing their number is not solved in general. One known case is that of the alphabet whose letters are cluster coordinates on $\text{Gr}(2,n)$ for which the integrable symbols of arbitrary weight have been enumerated in \cite{Brown:2009qja}. 

Our goal in this paper is introduce the \textsf{Mathematica} package \textsf{SymBuild} whose job it is to compute high-weight integrable symbols rapidly and to then manipulate them (for the purpose of computing derivatives or limits of them) efficiently. In order to solve the large matrix equations that arise at high weight, we use the \textsf{C++} linear algebra package \text{SpaSM} \cite{spasm}, which allows for efficient parallelization \footnote{The Mathematica interface for \text{SpaSM} and the finite field lift code can be downloaded from \url{https://bitbucket.org/yzhphy/linear_algebra_tools/src/master/}. (Please follow the readme.txt therein for its installation.) This interface also apply for IBP and differential equation computations in scattering amplitudes.}.

The outline of the paper is as follows.
In section~\ref{sec:notation} we introduce the notation and necessary definitions for the problem. 
In section~\ref{sec:organization} we present the organization of the \textsf{Mathematica} package and then explain the workings and functionalities of the various algorithms used. In section~\ref{sec:common commands}, we present the most important commands of the package and comment on their usage. 
We conclude in section~\ref{sec:conclusions}.

The package \textsf{SymBuild}, version 1.0, can be downloaded from,
\begin{quote}
\url{https://github.com/vladimirmitev/SymbolBuilding-development/archive/v1.0.tar.gz}
\end{quote}
The package file itself is contained in the file \textsf{SymBuild.wl}. The other file, the notebook \textsf{PackageTesting.nb}, contains examples of the different functionalities of the package. The other directories are not important for the end users.

\section{Setup and notation}
\label{sec:notation}

We start with an alphabet $\mathbb{A}=\{W_1,\ldots, W_M\}$ of algebraic functions (called \textbf{letters}) $W_i(\mathbf{x})$ in the kinematic variables $\mathbf{x}=(x_1,\ldots, x_N)$. In particular, the functions $W_i$ can contain roots $\rho_r(\mathbf{x})$, which we will deal with in section~\ref{subsec: dealing with roots}. We demand that the letters be independent in the sense that the symbols $[W_{1}],\ldots, [W_M]$ are \textbf{linearly independent.} This means that
\beq
\label{eq: alphabet independence}
\sum_{i=1}^M\alpha_i\, d \log(W_i)=\sum_{j=1}^Ndx_j\left(\sum_{i=1}^M\alpha_i\frac{\partial \log(W_i)}{\partial x_j}\right)=0\ \Rightarrow \alpha_i=0 \ \forall i\,.
\eeq
Let $|\mathbb{A}|$ be the number of elements in an alphabet, here $M$. We remark that two alphabets $\mathbb{A}=\{W_i\}$ and $\mathbb{A}={W_i'}$ are equivalent if there exists an invertible matrix with rational entries $M_{ij}$ s.t. $d\log W_i= \sum_j M_{ij}d\log W_j'$. 

Let us introduce the set of \textbf{iterated integrals} in alphabet $\mathbb{A}$. Such an integral (of \textbf{weight} $L$) is defined by picking a path $\gamma :[0,1]\rightarrow \mathbb{C}^N$ and computing the ordered iterated contour integral (beware the change of order between the LHS and the RHS)
\beq
\label{eq: definition contour integral of weight L}
\begin{split}
[W_{i_1},\ldots, W_{i_L}]_\gamma&\,=\,\int_\gamma d\log W_{i_L}\int_\gamma d\log W_{i_{L_1}}\cdots \int_\gamma d\log W_{i_1}\\
&=\,\int_{0}^1dt_{L}\dot{\gamma}_{r_L}(t_L)\partial_{r_L}W_{i_L}(\gamma(t_L))\int_{0}^{t_L}dt_{L-1}\dot{\gamma}_{r_{L-1}}(t_{L-1})\partial_{r_{L-1}}W_{i_{L-1}}(\gamma(t_{L-1}))\\&\quad \cdots \int_{0}^{t_2}dt_{1}\dot{\gamma}_{r_1}(t_1)\partial_{r_1}W_{i_1}(\gamma(t_1))\,,
\end{split}
\eeq
where $\partial_s\equiv \frac{\partial}{\partial x_s}$, $\dot{\gamma}(t)\equiv \partial_t\gamma(t)$ and we use the convention of summing over repeated indices. Let us denote by $x_b$ the initial or boundary point of the path, $x_b=\gamma(0)$ and simply by $x$ the final point $x=\gamma(1)$. We can generalize the set of iterated integrals that we look at by taking linear combinations of $\eqref{eq: definition contour integral of weight L}$ of the form $\sum_{I} c_{I} [W_{i_1},\ldots, W_{i_L}]_\gamma$, where $I=(i_1,\ldots, i_N)$ is a multiindex and the $c_I$ are constants. Hence, we consider the set of iterated integrals of weight $L$ as a vector space over the rational numbers. Furthermore, the space of all iterated integrals is a graded algebra, since the product of two iterated integrals of respective weights $L_1$ and $L_2$ is an iterated integral of weight $L_1 + L_2$.

In general, such iterated integrals will depend not just on the initial and final point, but also on the specific path $\gamma$. We can make a specific linear combination independent of \textbf{infinitesimal} variations of the contour by carefully choosing the constants such that they satisfy the relations
\beq
\label{eq: integrability condition}
\sum_{I} c_{I} [W_{i_1},\cdots, \hat{W}_{i_r},\hat{W}_{i_{r+1}},\cdots, W_{i_L}]_{\gamma}\Big(\partial_a \log W_{i_r}\partial_b \log W_{i_{r+1}}-(a\leftrightarrow b)\Big)=0
\eeq
for all $r=1,\ldots, L-1$ and for all $a<b=1,\ldots, N$.  Iterated integrals that satisfy \eqref{eq: integrability condition} depend only on the initial and final point of the path as well as on its homotopy class.

Having introduced iterated integrals, we shall now move on to discuss their \textbf{symbols}. First, a note on notation: we denote the symbol of a function $f$ as $SB[f]$. In particular, the symbol of the iterated integrals of \eqref{eq: definition contour integral of weight L} is written as\footnote{The following notations are equivalent: $[W_1,\ldots, W_n]\equiv [W_1]\otimes \cdots \otimes [W_n]$.}
\beq
\SB\left[\sum_{I}c_I [W_{i_1},\ldots, W_{i_L}]_{\gamma} \right]= \sum_{I}c_I [W_{i_1},\ldots, W_{i_L}]\,.
\eeq
Having introduced the notation for the symbols, let us talk about the symbols themselves. The symbol $\SB[f]$ is essentially a generalized primitive (i.e. a solution of a differential equation) and contains all the information on the positions of the branch points and singularities of the function $f$ as well as of its derivatives. We shall refrain from presenting a fully self-contained introduction to symbols here and refer to the article \cite{Duhr:2011zq} instead. The important pieces of information that the reader needs is that, due to the properties of the logarithm, the following identities hold for the symbols:
\beq
\begin{split}
[\cdots, A B,\cdots]&\,=\,[\cdots, A,\cdots]+[\cdots, B,\cdots]\,,\qquad 
[\cdots, A^{-1},\cdots]\,=-\,[\cdots, A,\cdots]\,,\\
[\cdots, \text{const},\cdots]&\,=\,0\,.
\end{split}
\eeq
Furthermore, if one is armed with the symbol of a function and knowledge of the appropriate boundary conditions, it becomes possible to reconstruct the function itself \cite{Goncharov:2010jf,Duhr:2011zq}. 
As an example of a symbol of a very well known function, let us take the dilogarithm. 
Since $\text{Li}_2(x)=-\int_{0}^xd\log(x')\int_{0}^{x'}d\log(1-x'')$, we obtain $\SB[\text{Li}_2(x)]=-[1-x,x]$. Finally, we remark that the first entry $W_{a_1}$ of the symbol $[W_{a_1},\ldots]$ contains information on the its discontinuities, while derivatives in act on the last entry; for instance $\partial_x \SB[\text{Li}_2(x)]$ is given by $-[1-x]\partial_x \log(x)=-\tfrac{1}{x}[1-x]$. 

The main goal of this article is to provide a tool for the computation and manipulation of all \textbf{integrable symbols} of arbitrary weight $L$ in a given alphabet $\mathbb{A}=\{W_1,\ldots, W_M\}$, i.e. of all linear combination $\sum_{I}c_I [W_{i_1},\ldots, W_{i_L}]$ that satisfy the condition \eqref{eq: integrability condition} (with the $\gamma$ subscript in \eqref{eq: integrability condition} dropped). In order to systematize this task, let us make some further definitions. Let $\calV_L$ be the space of \textbf{all} weight $L$ symbols
\beq
\calV_L\,=\,\text{span}\big\{[W_{a_1},\ldots , W_{a_L}]: a_i\in\{1,\ldots, M\}\big\}\,,
\eeq
and set $\calV\equiv \calV_1$ and  $\calV_0=\mathbb{Q}$. Clearly $\calV_L\cong \calV^{\otimes L}$ so we can talk about both interchangeably. Let $\calH_L$ be the subset of \textbf{integrable} symbols, i.e. the subspace of $\calV_L$ of linear combinations that satisfy \eqref{eq: definition contour integral of weight L}. We can also impose first-, second-, ..., last-entry conditions on $\calH_L$. For us, an $n-$entry condition is a set $\mathcal{C}_{n}=\{s^i\}$ of \textbf{forbidden} sequences $s^{i}=\{s^i_1,\ldots, s^i_n\}$ that excludes symbols of the type $[W_{s^{i}_1},\ldots, W_{s^{i}_n},\star]$ from appearing in $\calH_L$. These conditions are independent of the integrability conditions.

\subsection{The integrability tensor}

In order to deal with the integrability condition \eqref{eq: integrability condition} in a more systematic way, we define the following set of algebraic functions
\beq
F_{ij}^{(rs)}(x)\,=\, \frac{\partial \log W_i}{\partial x_r}\frac{\partial \log W_j}{\partial x_s}-\big(r\leftrightarrow s\big)\,.
\eeq
Due to the antisymmetry, we can restrict to $r<s=1\,\ldots, N$ and $i<j=1,\ldots, M$. 
First, we are interested in the set of linearly independent constant solutions $c_{ij}$ to the set of equations
\beq
\label{eq: initial integrability conditions}
\sum_{i<j=1}^M c_{ij}F_{ij}^{(rs)}(x)=0\,,\qquad \forall x\,\text{ and }\, \forall r,s\,.
\eeq
Instead of the set of solutions, we are actually more interested in the set of constraints. That is, we need to determine a \textbf{constant} tensor $\mathbb{F}_{ij}^A$ with $i,j=1,\ldots, M$ and $A=1,\ldots, R$ that is antisymmetric ($\mathbb{F}_{ij}^A=-\mathbb{F}_{ji}^A$), such that the solution space to the set of equations $\sum_{i<j=1}^Mc_{ij}\mathbb{F}_{ij}^A=0$ for all $A$ is the same as the space of solutions to \eqref{eq: initial integrability conditions}. In this setup, the number $R$, the integrability rank of the alphabet $\mathbb{A}$, is the minimal number for which this is possible. We shall refer to $\mathbb{F}_{ij}^A$ as the \textbf{integrability tensor} and we can represent it as a list (of length $R$) of $M\times M$ antisymmetric matrices. As an example, consider the alphabet $\mathbb{A}=\{x_1,x_1-x_2,x_2\}$ for which
\beq
\{F_{12}^{(12)},F_{13}^{(12)},F_{23}^{(12)}\}=\left\{-\frac{1}{x_1 (x_1-x_2)},\frac{1}{x_1 x_2},\frac{1}{x_2 (x_1-x_2)}\right\}\,.
\eeq
It is easy to see that the three functions satisfy a single linear equation and thus $R=2$. We find 
\beq
\mathbb{F}^1=
\left(
\begin{array}{ccc}
 0 & 1 & 0 \\
 -1 & 0 & -1 \\
 0 & 1 & 0 \\
\end{array}
\right)\,,
\qquad 
\mathbb{F}^2=\left(
\begin{array}{ccc}
 0 & 0 & 1 \\
 0 & 0 & 1 \\
 -1 & -1 & 0 \\
\end{array}
\right)\,.
\eeq
The advantages of working with the integrability tensor is that 1) it is constant and 2) it is minimal and packs in a neat way all the conditions coming from the integrability equation \eqref{eq: integrability condition}. Once it has been determined, the computation of the integrable symbols can be done without any reference to the functional form of the letter of the alphabet $\mathbb{A}$.

\subsection{Iterative notation}

To each weight $L$ integrable symbol $S^{(L)}\in \calH_L$ corresponds by definition a tensor $c_I$ with $L$ indices $S^{(L)}=\sum_I c_I [W_{i_1},\ldots, W_{i_L}]$. Such linear combinations can easily end up having millions of terms and thus be prohibitively demanding of computer memory as $L$ gets large. It is therefore necessary to use an alternate method of representation.  In the \textbf{iterative notation}, introduced by Dixon et al. \cite{Dixon:2013eka}, the symbols are written as\footnote{We let the index $j_L$ label the basis elements as $\calH_L=\text{span}\{S_{j_L}^{(L)}\}_{j_L=1}^{\dim \calH_L}$.}
\beq
\label{eq: representing the symbols iteratively}
S^{(L)}_{j_L}=\sum_{j_{L-1}=1}^{\dim \calH_{L-1}}\sum_{i_L=1}^Md_{j_L}^{j_{L-1}i_L}S^{(L-1)}_{j_{L-1}}\otimes [W_{i_L}]\,.
\eeq
We shall refer to the 3-index tensors $d_{j_L}^{j_{L-1}i_L}$ as the weight $L$ \textbf{integrable tensors}. The set of integrable tensor $\{d_{j_1}^{j_{0}i_1},d_{j_2}^{j_{1}i_2},\cdots, d_{j_L}^{j_{L-1}i_L}\}$ contains all the information needed to work with the integrable symbols of weight $L$. In all cases we have worked out, the integrable tensors are very sparse, boasting a density of less than one percent.

Iterating the notation \eqref{eq: representing the symbols iteratively} one step further, and using the fact that by definition $\calH_0=\mathbb{Q}$, so that $j_0=1$ is a trivial index, we obtain 
\beq
S^{(L)}_{j_L}=\sum_{i_1,\ldots, i_L=1}^M\underbrace{\sum_{j_1,\ldots, j_{L-1}}d_{j_1}^{j_0i_1}d_{j_2}^{j_1i_2}\cdots d_{j_L}^{j_{L-1}i_L}}_{=(c_{j_L})_{i_1,\ldots, i_L}}[W_{i_1},\ldots, W_{i_l}]\,.
\eeq
One of the advantages of the iterative notation is the fact that the integrability condition \eqref{eq: integrability condition} for the weight $L$ symbols (assuming that the symbols of lower weight are integrable) can be written in a very simple manner:
\beq
\label{eq: iterative weight L integrability condition}
\begin{split}
&\sum_{j_{L-1}=1}^{\dim \calH_{L-1}}\sum_{i_{L-1},i_L=1}^M d_{j_L}^{j_{L-1}i_L}d_{j_{L-1}}^{j_{L-2}i_{L-1}}\mathbb{F}_{i_{L-1},i_{L}}^A=0\,,\\
&\qquad \forall A=1,\ldots, R\text{ and }\forall j_{L-2}=1,\ldots, \dim \calH_{L-2}\,.
\end{split}
\eeq
The above can be rewritten as a matrix equation with $(\dim H_{L-2})\times R$ rows and $(\dim H_{L-1})\times M$ columns for the unknowns $d_{j_L}^{j_{L-1}i_L}$. We assume implicitly that the weight $L-1$ integrable tensor $d_{j_{L-1}}^{j_{L-2}i_{L-1}}$ has already been determined.  We remind that the index $j_L$ in \eqref{eq: iterative weight L integrability condition} is a dummy index that simply enumerates the elements of a basis set of solutions. As it turns out, in all cases tested, \eqref{eq: iterative weight L integrability condition} is a sparse system of equations that can be solved very efficiently with \textsf{SpaSM} or directly in \textsf{Mathematica} if the size of the system is not too big. We remark that it is also common and possible to require that the last entries of the integrable symbols come from some other (smaller) set of letters $\mathbb{A}'$ whose members can be expressed as linear combinations of the letters of $\mathbb{A}$. Implementing such conditions in straighforward in this framework and the whole procedure is implemented in the command \textsf{determineNextWeightSymbols}, see section~\ref{sec:common commands} for more information.

\subsection{Counting products}

One interesting question appears often: what is the number of integrable symbols of weight $L$ that are products? The product of a function $f_1$ of weight $L_1$ and of a function $f_2$ of weight $L_2$ is a function of weight $L_1+L_2$ and its symbol is given by the shuffle product of the symbols $\SB[f_1]$ and $\SB[f_2]$: 
$\SB[f_1 f_2]=\SB[f_1]\shuffle \SB[f_2]$, where 
\beq
[W_{i_1},\ldots, W_{i_{L_1}}]\shuffle [W_{j_1},\ldots, W_{j_{L_2}}]=\sum_{\sigma\in S(L_1,L_2)}[W_{k_{\sigma(1)}},\ldots, W_{k_{\sigma(L_1+L_2)}}]\,,
\eeq
where $\{k_1,\ldots, k_{L_1+L_2}\}=\{i_1,\ldots, i_{L_1},j_1,\ldots, j_{L_2}\}$ and 
\beq
S(L_1,L_2)=\left\{\sigma\in S_{L_1+L_2}: \begin{array}{l}\sigma^{-1}(1)<\sigma^{-1}(2)<\cdots<\sigma^{-1}(L_1)\\ \sigma^{-1}(L_1+1)<\sigma^{-1}(L_1+2)<\cdots<\sigma^{-1}(L_1+L_2) \end{array}\right\}\,.
\eeq
The shuffle product of two integrable symbols is again integrable. 

We define the \textbf{product symbols} at weight $L$ to be the integrable symbols that are obtained by using the shuffle product on lower weight symbols and we call the associated space $\calP_L$. The space of \textbf{irreducible} symbols $\calQ_L$ is then defined as
\beq
\calQ_L=\calH_L/\calP_L\,.
\eeq
In particular $\calQ_1=\calH_1$. We find via simple combinatorics that
\beq
\label{eq: dimensions of the product symbols}
\dim \calP_L=\sum_{\substack{\lambda =\{n_1,\ldots, n_S\}\\\ n_i\in \mathbb{N}\,,\ S>1\\ \sum_{i=1}^Si n_i=L}}\prod_{j=1}^S\binom{\dim\calQ_j+n_j-1}{n_j}\,,
\eeq 
i.e. it is a sum over partitions $\lambda$ and the binomials take care of the appropriate symmetrizations. Using \eqref{eq: dimensions of the product symbols} and the obvious equation $\dim\calH_L=\dim\calP_L+\dim\calQ_L$, we can easily express $\dim\calQ_L$ through $\dim\calH_i$. The counting of symbols that are products/are irreducible is implemented in the package \textsf{SymBuild} using the commands \textsf{dimProductSymbols}/\textsf{dimIrreducibleSymbols}, see the auxiliary example file.

As a corrolary of \eqref{eq: dimensions of the product symbols}, it follows that we have the following dimensions bounds (ignoring $n$-entry conditions)
\beq
2^M\geq \dim \calH_L\geq \binom{M+L-1}{L}\,,
\eeq
where the right bound is saturated for the alphabet $\mathbb{A}=\{x_1,\ldots, x_N\}$. The space of functions for that alphabet is made up only of products of logs.  

In addition to computing the dimension of the space of irreducible symbols, we can project onto it explicitly. Specifically, product symbols are annihilated by the $\rho$ operation, see for example \cite{Golden:2014xqa}. This map $\rho:\calV_L\rightarrow \calV_L$ is defined recursively by setting $\rho([W])=[W]$ and then demanding that
\beq
\rho([W_1,\ldots, W_k])=\frac{k-1}{k}\Big[\rho([W_1,\ldots, W_{k-1}])\otimes [W_k]-\rho([W_2,\ldots, W_{k}])\otimes [W_1]\Big]\,,
\eeq
for $k>1$. The map $\rho$ is implemented in the package by the command \textsf{removeProductsFromSymbolTensorArray}, see the auxiliary example file.
In priniple, we have the identifications $\calP_L=\mker(\rho)_{\calH_L}$ and  $\calQ_L=\mim(\rho)_{\calH_L}$. However, if we have imposed some special $n$-entry conditions, it can be the case that not all elements annihilated by $\rho$ can be written as products of lower weight symbols.

\section{Explanation of the algorithms}
\label{sec:organization}

\begin{figure}[htbp!]
             \begin{center}       
              \includegraphics[scale=0.45]{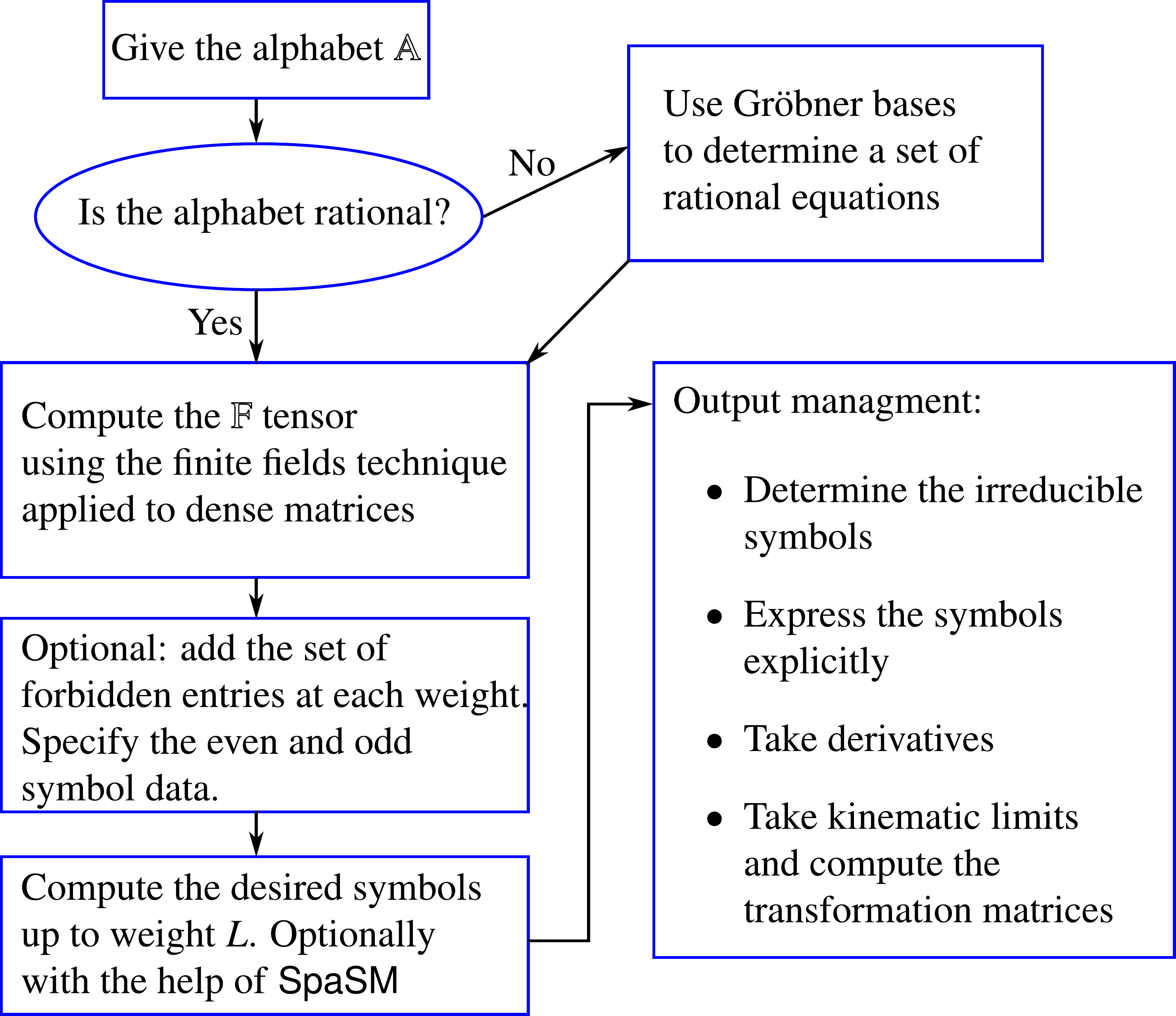}
              \caption{Flowchart describing the main steps in computing the integrable symbols as well as the possible operation that can be applied to the results. }
              \label{fig:flowchart}
            \end{center}
\end{figure}

In this section, we present the organization of the package as well as the inner working of the various algorithms. We begin with the steps to follow when computing the symbols, shown in figure~\ref{fig:flowchart}.

\subsection{Computing the integrability tensor}
\label{subsec: Computing the integrability tensor}

The computation of the $\mathbb{F}^A_{ij}$ tensor proceeds as follows. If the alphabet $\mathbb{A}$ is rational, then the set of equations \eqref{eq: initial integrability conditions} involves only rational functions in the variables $x_r$. If the alphabet is not rational, then we need to use Gr\"obner bases first in order to obtain a set of equations involving purely rational functions. This will be discussed in section~\ref{subsec: dealing with roots}. Ignoring that case for now, we shall generate the integrability tensor from \eqref{eq: initial integrability conditions} in the following way:
We start with an empty matrix $\mathbb{M}$
\begin{enumerate}
\item We first take one of the equations of \eqref{eq: initial integrability conditions}, say for $r=1$ and $s=2$. 
\item Set the variables $x_r$ to random values. For this, we choose random primes numbers (determined by the parameter \textsf{sampleSize}) such  that the $x_r$ are all different. This gives a row of $\binom{M}{2}$ constants that we append to $\mathbb{M}$. 
\item We repeat this \textsf{maxSamplePoints} number of times.  We go back to step 1. and do the same for the next equation in  \eqref{eq: initial integrability conditions} until we run out of equations.
\item We can now row reduce the $\mathbb{M}$ matrix and drop the zero rows at the end. The final rank of $\mathbb{M}$ is the integrability rank $R$.
\end{enumerate}
We can then directly extract $\mathbb{F}^{A}_{ij}$ from $\mathbb{M}_{A I}$ by unrolling the indices appropriately. In all cases that were tested, the integrability tensor $\mathbb{F}^{A}_{ij}$ is a very sparse tensor (less than 1\% of occupancy) and its entries tend to be simple rational numbers.
All along, the computations are done in a finite field $\mathbb{Z}_p$ for a given large prime number $p$, after which we apply Wang's reconstruction algorithm. The procedure is then repeated for a different large prime $p'$ and if the results agree, the computation terminates.

\subsection{Dealing with roots}
\label{subsec: dealing with roots}

The alphabet $\mathbb{A}$ in the variables $\mathbf{x}=\{x_1,\ldots, x_N\}$ can contain a number $Q$ of roots $\boldsymbol{\rho}=\{\rho_1,\ldots, \rho_Q\}$, which are of course functions in the variables $x_i$. We shall assume that the roots are zeroes of the $Q$ minimal polynomials $\{R_i(\mathbf{x},\boldsymbol{\rho})\}_{i=1}^Q$, where $R_i$ is of degree $m_i$. The idea is to rewrite every $F_{ij}^{(rs)}(\mathbf{x},\boldsymbol{\rho})$ in \eqref{eq: initial integrability conditions}, which is a rational function of the $x$'s and of the $\rho$'s as 
\beq
F_{ij}^{(rs)}(\mathbf{x},\boldsymbol{\rho})=\sum_a\sum_{n_a=0}^{m_a-1}\underbrace{f_{ij}^{(rs)}(n_1,\ldots, n_Q; \mathbf{x})}_{\text{rational function}} \rho_1^{n_1}\cdots \rho_Q^{n_Q}\,.
\eeq
Since the roots are algebraically independent, each integrability equation \eqref{eq: initial integrability conditions} (for a given fixed $r$ and $s$) gives now rise to $m_1\times  \cdots \times m_Q$ equations involving only the purely rational functions $f_{ij}^{(rs)}(n_1,\ldots, n_Q; \mathbf{x})$. This allows one to apply the same procedure as for the rational alphabets and to thus determine the integrability tensor for that alphabet.

\subsection{Computing the integrable symbols}

Once the integrability tensor has been computed, we can proceed to solve \eqref{eq: iterative weight L integrability condition} iteratively, starting from $L=1$ (for which there is nothing to solve) and continuing to a used specified maximal weight $L_{\text{max}}$. Before the computation is done, we may optionally specify
\begin{enumerate}
\item the $n-$entry conditions, i.e. which letters, or combinations of letters are forbidden from appearing in the $n$ entry of the symbols. 
\item which letters of the alphabet are even and which are odd. The package \textsf{SymBuild} then automatically separates the solutions accordingly in even and odd symbols. In this context even/odd refers to the behavior of the letters $W_i$ under an automorphism $\pi$ (such as for example complex conjugation) under which $d\log W_i\rightarrow \pm d\log W_i$ with $+$ for even and $-$ for odd. 
\end{enumerate}

Depending on the options specified, one may use  \textsf{determineNextWeightSymbolsSimple} or \textsf{determineNextWeightSymbolsSimple} as explained more thoroughly in section~\ref{sec:common commands}. Thus, the computation of the sets of integrable symbols proceeds iteratively. We would like to remark here that an alternative algorithm for the computation of weight $L$ integrable symbols, in which the the equations to solve arise from consistency conditions for the gluing of weight $a$ and weight $L-a$ symbols, was presented in \cite{Drummond:2014ffa}. While it has some advantages, we preferred for simplicity to implement the standard iterative method here in which the weight $L$ integrable symbols are constructed from the already computed weight $L-1$ ones.

\subsection{The inversion tensor and transformations of symbols}
\label{subsec: inversion tensor}

The weight $L$ \textbf{inversion tensor} $E_{j_{L-1}i_L}^{j_L}$ that has the following property. Given an arbitrary integrable tensor of weight $L$, written iteratively as $S^{(L)}=\sum_{j_{L-1}=1}^{\dim \calH_{L-1}}\sum_{i_L=1}^Mc^{j_{L-1}i_L}S_{j_{L-1}}^{(L-1)}\otimes[W_{i_L}]$, then that symbol can be expanded in the basis of $\calH_{L}$ given by the integrable tensors \eqref{eq: representing the symbols iteratively} as  
\beq
S^{(L)}=\sum_{j_L=1}^{\dim \calH_{L}}\left(\sum_{j_{L-1}=1}^{\dim \calH_{L-1}}\sum_{i_L=1}^Mc^{j_{L-1}i_L}E_{j_{L-1}i_L}^{j_L}\right)S^{(L)}_{j_L}\,.
\eeq
The inversion tensor $E_{j_{L-1}i_L}^{j_L}$ is the left-inverse of the integrable tensor $d^{j_{L-1}i_L}_{j_L}$ in the following sense: we can think of the pair $(j_{L-1},i_L)$ as a single index $I=|\mathbb{A}|(j_{L-1}-1)+i_L \in \{1,\ldots, |\mathbb{A}|\dim\calH_{L-1}\}$. Applying this procedure on the integrable tensor $d^{j_{L-1}i_L}_{j_L}$, we obtain a $|\mathbb{A}|\dim\calH_{L-1}\times \dim\calH_{L}$ matrix $d^{I}_{j_L}$ that has more rows than columns and has rank $\dim\calH_{L}$ by construction. Such a matrix has a (non-unique) left inverse matrix $E_{I}^{j_L}$ (with $E_{I}^{j_L}d^{I}_{j_L'}=\delta_{j_L'}^{j_L}$) for which we can replace $I$ by the bi-index $(j_{L-1},i_L)$ to obtain the desired 3-tensor.

The inversion tensor is useful in determining the transformation matrices implementing various symmetries of the alphabet on the space of integrable symbols, as well as in computing the limits of integrable symbols. Specifically, if one is given a linear transformation $T$ mapping the weight 1 symbols in one alphabet $\mathbb{A}$ to those of another\footnote{Of course, nothing forbids the case $\mathbb{A}'=\mathbb{A}$.} $\mathbb{A}'$ 
\beq
T([W_i])=\sum_{i'=1}^{|\mathbb{A}'|}\mathbb{T}_{i}^{\phantom{i}i'}[W'_{i'}]\,,
\eeq
then we can compute the transformation map $T:\calH_L\rightarrow \calH_{L}'$ acting on arbitrary integrable symbols iteratively as
\beq
\label{eq: transformation matrix of weight L}
T(S_{j_L}^{(L)})\equiv{T^{(L)}}_{j_L}^{\phantom{j_L}j'_L}{S'_{j'_L}}^{(L)}=\left(d^{j_{L-1}i_L}_{j_L}{E'}_{j'_{L-1}i'_L}^{j'_L}\mathbb{T}_{i_L}^{\phantom{i_L}i'_L}{T^{(L-1)}}_{j_{L-1}}^{\phantom{j_{L-1}}j'_{L-1}}\right){S'_{j'_L}}^{(L)}\,,
\eeq
where we have summed over repeating indices. Thus, we have determined the weight $L$ transformation map in terms of the weight $L$ inversion tensor ${E'}_{j'_{L-1}i'_L}^{j'_L}$ for the $\mathbb{A}'$ symbols, the weight $L$ $\mathbb{A}$ integrable tensor $d^{j_{L-1}i_L}_{j_L}$, the weight $L-1$ transformation matrix ${T^{(L-1)}}_{j_{L-1}}^{\phantom{j_{L-1}}}$ and the original transformation input $\mathbb{T}_{i}^{\phantom{i}i'}$.  Since at weight zero, the transformation matrix trivially ${T^{(0)}}_{j_0}^{\phantom{j_0}j'_0}={T^{(0)}}_{1}^{\phantom{1}1}=1$, we can directly compute the other transformation matrices. 


\section{Most common commands}
\label{sec:common commands}

This article is accompanied by an annotated \textsf{Mathematica} file that shows explicitly how to compute the integrable symbols of a variety of alphabets (with or without roots). It further contains examples of implementing symmetry and limit operations. In this section, we want to emphasize the most important parameters and commands of the package. 

\begin{itemize}
\item \textsf{computeTheIntegrabilityTensor} takes as inputs an alphabet (array of functions), the array of variables of the alphabet, the list of roots appearing in the alphabet (is an empty list if the alphabet is rational), the list of minimal polynomials describing the roots, an optional list of replacement rules for voluminous objects in the list of minimal polynomials, and three further integers. Those integers are: 1) one determining the size of the random prime numbers used in the algorithm of section~\ref{subsec: Computing the integrability tensor}, 2) one setting the number of sampling points and 3) the last one specifying the number of retries allowed if the sample points turn out to contain infinities. 
\item \textsf{findRelationsInAlphabet} takes the same inputs as \textsf{computeTheIntegrabilityTensor} and computes all linear relations existing between the letters of the alphabet in the sense of \eqref{eq: alphabet independence}. 
\item \textsf{weight1Solution} and \textsf{weight1SolutionEvenAndOdd} are two commands that prepare the weight 1 integrable tensors. The first command takes an alphabet and a list (potentially empty) of forbidden first entries, while the second also takes as second argument a list of digits (0 for even, 1 for odd) that indicate which letters are even/odd. The first command returns an the weight 1 integrable tensor, while the second returns an array of two elements consisting of the weight 1 integrable tensor and an array indicating their signs.
\item \textsf{determineNextWeightSymbolsSimple} and \textsf{determineNextWeightSymbols} are two commands that compute the next weight symbols. The simplified command \textsf{determineNextWeightSymbolsSimple} takes the weight $L-1$ integrable tensor and the integrability tensor $\mathbb{F}$ (and two optional parameters, see the auxiliary example file) and directly determines the weight $L$ integrable tensor. On the other hand, the command \textsf{determineNextWeightSymbols} takes (in this order) the weight $L-1$ integrable tensor, the list of its signs, the integrability tensor, the list determining the signs of the letters, an optional matrix of equations that determine the forbidden entries at weight $L$ (either $=\textsf{False}$ or a matrix generated with the command \text{weightLForbiddenSequencesEquationMatrix}) and an optional matrix that gives the last entries of the symbol as linear combinations of the letters of the alphabet. This more complicated command returns the integrable tensor of weight $L$ as well as a list indicating the signs of the integrable symbols.
\item \textsf{weightLForbiddenSequencesEquationMatrix} is a command that takes the weight $L-1$ integrable tensor, a list of forbidden sequences  of entries at weight $L$ and the size of the alphabet as inputs and then generates a matrix of equations that is used in \textsf{determineNextWeightSymbols} to remove the forbidden weight $L$ entries.
\item \textsf{presentIntegrableSymbolsData} is a command that takes either an integrable tensor or a list consisting of an integrable tensor and the array indicating its signs, and presents them in a legible way. 
\item \textsf{expressTensorAsSymbols} takes a product of integrable tensors and presents them as a list of explicit formal symbols in several possible ways. Specifically, its inputs are a product
\beq
\textsf{d}(j_{a-1};i_a,\ldots, i_b;j_b)\equiv d_{j_a}^{j_{a-1}i_a}d_{j_{a+1}}^{j_{a}i_{a+1}}\cdots d_{j_{b}}^{j_{b-1}i_{b}}
\eeq
where repeated indices are summed over ($a\leq b$) and an optional string \text{opt} giving the display specification. The object $\textsf{d}$ is a tensor with $3+b-a$ indices that contains all the information for the integrable symbols between weight $a$ and $b$. Plugging $\textsf{d}$ into \textsf{expressTensorAsSymbols} together with \text{opt}=``Recursive" leads to the output
\beq
\label{eq: explicit symbol expression}
\left\{\sum_{I}c_{j_{a-1};I}^{(1)}\text{sb}[\mathcal{S}[j_{a-1}],\{i_{a},\ldots, i_b\}], \sum_{I}c_{j_{a-1};I}^{(2)}\text{sb}[\mathcal{S}[j_{a-1}],\{i_{a},\ldots, i_b\}]\ldots \right\}
\eeq
which is an array of length $\dim\calH_b$ of explicit integrable symbols. In the above, $\mathcal{S}$ are placeholders for the integrable symbols of weight $a-1$. Furthermore, if $a=1$ and the option \text{opt}=``Complete" is used, then the $\mathcal{S}$ part of the expression \eqref{eq: explicit symbol expression} is dropped since there is only one symbols of weight 0. 
\item \textsf{convertFormalSymbol} takes two inputs: 1) an integrable symbols written as $\sum_{I} c_I \text{sb}[\{i_1,\ldots, i_L\}]$ and 2) an alphabet $\mathbb{A}=\{W_1,W_2,\ldots\}$. Its output is an expression of \textbf{explicit symbols} $\sum_{I} c_I \text{SB}[\{W_{i_1},\ldots, W_{i_L}\}]$, which can sometimes be useful for the computation of limits of symbols. 
\item \textsf{symbolDerivative} takes three elements as input: 1) a symbol (written like in \eqref{eq: explicit symbol expression}) whose derivative is to be computed, 2) the alphabet in which the symbol is written and 3) the variable w.~r.~t.~to which to take the derivative. Alternatively, it can act on expressions of explicit symbols SB (such as those produces by \textsf{convertFormalSymbol}) in which case the alphabet should not be given. 
\item \textsf{IntegrableQ} is a command that can be applied to a linear combination $S$ of formal symbols (either sb or SB) to determine if $S$ is integrable. If acting on sb symbols, the integrability tensor $\textsf{F}$ should be given as second input. If acting on the explicit SB symbols, the list of variables must be given as second input.
\item \textsf{determineLeftInverse} acts on a matrix $A$ of size $m\times n$ with $m\geq n$ but with full rank ($\text{rank}(A)=n$) and computes its left-inverse, i.e. a $n\times m$ matrix $B$ such that $BA=\mathbbm{1}_n$. 
\item \textsf{computeTheInversionTensor} takes a symbol tensor $d_{j_L}^{j_{L-1}i_L}$ and computes the inversion tensor $E_{j_{L-1}i_L}^{j_L}$ as explained in section~\ref{subsec: inversion tensor}.
\item \textsf{buildTransformationMatrix} computes the weight $L$ transformation matrix between the sets of integrable symbols in the alphabet $\mathbb{A}$ and those in the alphabet $\mathbb{A}'$, as explained in \eqref{eq: transformation matrix of weight L}. It takes (in this order) as input the weight $L$ integrable tensor $d_{j_{L}}^{j_{L-1}i_{L}}$ for $\mathbb{A}$, the transformation matrix $T^{(L-1)}$, the initial transformation matrix $\mathbb{T}$ and the weight $L$ inversion tensor $E_{j'_{L-1}i'_L}^{j'_L}$ for the alphabet $\mathbb{A}'$.
\item \textsf{radicalRefine} is an auxiliary command that takes a list of expressions involving roots and attempts to provide a system of minimal polynomials that can then be used in, for example, the command \text{computeTheIntegrabilityTensor}. For instance, $\{\text{mp}, \text{rules}\}= \textsf{radicalRefine}\left[\left\{\sqrt{a}+\sqrt{b},\sqrt{a}-\sqrt{b}\right\}\right]$ gives
\begin{align}
\text{mp}&=\left\{-a+b+\mathbb{X}(1) \mathbb{X}[2],-2 a-2 b+\mathbb{X}[1]^2+\mathbb{X}[2]^2,\mathbb{X}[2] (2 a+2 b)+\mathbb{X}[1] (b-a)-\mathbb{X}[2]^3\right\}\,,\nonumber\\
\text{rules}&=\left\{\sqrt{a}\to \frac{\mathbb{X}[1]}{2}+\frac{\mathbb{X}[2]}{2},\sqrt{b}\to \frac{\mathbb{X}[1]}{2}-\frac{\mathbb{X}[2]}{2}\right\}\,,
\end{align}
where ``mp" is the list of minimal polynomials in the auxiliary variables $\mathbb{X}[1]$ and $\mathbb{X}[2]$ and ``rules" tells us how to express the desired roots $\sqrt{a}$, $\sqrt{b}$ from the roots of the minimal polynomials \footnote{Any finite algebraic extension has a primitive element. In the usual sense, {\it the} minimal polynomial of an algebraic extension is a univariate polynomial defined for the primitive element. However, the minimal polynomial of a primitive element may involve complicated coefficients. Here ``minimal polynomials'' refer to the constraint multivariate polynomials for generators of an extension. Frequently, polynomials found by  \textsf{radicalRefine}  have simpler coefficients than these in the minimal polynomial of a primitive element. }.
\end{itemize}

All of these commands are illustrated in several examples in the \textsf{Mathematica} notebook accompanying this package. 

Furthermore, the package contains a number of global parameters that can be modified to adapt some of the algorithms used. These parameters are \textsf{globalLowerThreshold}, \textsf{globalSpaSMThreshold}, \textsf{globalGetNullSpaceStep}, \textsf{globalSpaSMListOfPrimes}, \textsf{globalGetNullSpaceSpaSMPrimes}, \textsf{globalSetOfBigPrimes}, \textsf{globalRowReduceOverPrimesInitialNumberOfIterations}, \textsf{globalRowReduceOverPrimesMaxNumberOfIterations}, \textsf{globalRowReduceOverPrimesMethod}, \textsf{globalRowReduceMatrixSpaSMPrimes} and \textsf{globalVerbose}. Their standard values can be seen at the beginning of the auxiliary example notebook and their purpose examined with the $?$ and $??$ commands. If not using \textsf{SpaSM}, the most important would be $\textsf{globalVerbose}$ (a boolean. Setting it to $\textsf{False}$ suppresses all monitoring messages) and \textsf{globalGetNullSpaceStep} which is the size of the pieces into which big matrices are cut when the program tries to determine their null space. Its default value is 200 and reducing that number can sometimes improve the performance for very large matrices.

\section{Applications and performance}
\label{sec:performance}

Let us now discuss the performance of the package. For this, we have timed the time needed for the computation of symbols of various weight in several alphabets. 

First, let us take the case of the 5-pt planar alphabet $\mathbb{A}_{\text{P}}$ of 26 letters, see for example \cite{Chicherin:2017dob} for a definition. On a recent Intel i7 laptop with 8Gb of RAM, the computation of integrability tensor and of the symbols of weight $L<6$ (subject to the first entry conditions allowing only the letters $\{W_i\}_{i=1}^5$ as first entries) takes in total about 500 seconds, see table~\ref{tab:number of integrable symbols and computation 1}. The situation for the non-planar alphabet $\mathbb{A}_{\text{NP}}$ of 31 letters is more complicated\footnote{See again  \cite{Chicherin:2017dob} for a definition and for an explanation of the first and second entry conditions.}. Even after imposing the second entry conditions that restrict the number of symbols significantly, it becomes very difficult, due to memory and patience constraints, to compute the number of integrable symbols of weight 5 on a laptop. Still, the remaining computation time is manageable, see again table~\ref{tab:number of integrable symbols and computation 1}.

\begin{table}[th]
\centering
\renewcommand{\arraystretch}{1.6}
\begin{tabular}{|l|c|c|c|c|c|c|c|}
\hline
& 
$\mathbb{F}$
&
1
&
2
&
3
&
4
&
5
&
6
\\
\hline
$\mathbb{A}_{\text{P}}$ (1$^{\text{st}}$ entry)& & 5$\,|\,$0 & 25$\,|\,$0 & 125$\,|\,$1 & 645$\,|\,$16 & 3275$\,|\,$161 & 17095$|$1331\\
computation time (s)& $ 25$& $<1$ & $<1$& $<2$ & $ 20$& $ 450$ (\textcolor{blue}{50}) & \textcolor{red}{37 hours} (\textcolor{blue}{7 hours})\\
$\mathbb{A}_{\text{NP}}$ ($1^{\text{st}}$ entry)&  & 10$\,|\,$0 & 100$\,|\,$9 & 1000$\,|\,$180 &  9946$\,|\,$2730 & &
\\
computation time (s)&$ 40$ &$<1$ & $ 1$& $ 18$ & $ 4700$ (\textcolor{blue}{1000}) & & \\
$\mathbb{A}_{\text{NP}}$ ($1^{\text{st}} \&$ $2^{\text{nd}}$ entry)&  & 10$\,|\,$0 & 70$\,|\,$9 & 505$\,|\,$111 & 3736$\,|\,$1191 & &\\
computation time (s)&$ 40$ &$<1$ & $ 1$& $ 12$ & $ 1150$ (\textcolor{blue}{240}) & &\\
\hline
\end{tabular}
\renewcommand{\arraystretch}{1.0}
\caption{ The number of independent integrable symbols of given weight for the planar and the non-planar 5-pt alphabets, together with the approximate computation time (in seconds unless noted otherwise, done on a system with quad i7 processors and 8Gb RAM) for the symbols and for the integrability tensor separately. In each case, we indicate the number of even$\,|\,$odd symbols, whose computation was also included in the computing time. These computations were done completely within Mathematica, without \textsf{SpaSM}. The computation time using \textsf{SpaSM} on the same machine is put in parenthesis and in blue. In red, we put the computations done on the computer cluster Mogon. We used just two cores, but a lot (300+ Gb) of RAM.}
\label{tab:number of integrable symbols and computation 1}
\end{table}

We have also tested the 6-point dual-superconformal alphabet $\mathbb{A}_{\text{H}}$ of 9 letters, see for example \cite{Dixon:2013eka} for a definition.  We find the number of symbols (with the respective computation time) in table~\ref{tab:number of integrable symbols and computation 2}. The computations take comparately longer than the similarly sized ones in table~\ref{tab:number of integrable symbols and computation 1} because the integrability equations are not as sparse.
\begin{table}[th]
\centering
\renewcommand{\arraystretch}{1.6}
\begin{tabular}{|l|c|c|c|c|c|c|c|}
\hline
& 
$\mathbb{F}$
&
3
&
4
&
5
&
6
&
7
&
8
\\
\hline
$\mathbb{A}_{\text{H}}$ & & 26& 75 & 218 & 643 & 1929 & 5897
\\
computation time (s)& $< 1$& $<1$ & $<1$& $~1$ & $ 6$& $ 90$ &   (\textcolor{blue}{13000})
\\
$\mathbb{A}_{\text{H}}$ (last entry)&  & 7 & 21 & 62 &  188 & 579 & 1821
\\
computation time (s)& $< 1$& $<1$ & $<1$& $<1$ & $ 4$& $ 45$ & (\textcolor{blue}{1000})
\\
\hline
\end{tabular}
\renewcommand{\arraystretch}{1.0}
\caption{ We list here the number of independent integrable symbols (note that these are not irreducible symbols!) of given weight for 6-point dual-superconformal alphabet of 9 letters, together with the approximate computation time  (i7 processor, 8Gb RAM) for the symbols and for the integrability tensor separately. These computations were done completely within Mathematica, without \textsf{SpaSM}. The computation time using \textsf{SpaSM} on the same machine is put in parenthesis and in blue. }
\label{tab:number of integrable symbols and computation 2}
\end{table}

\section{Conclusions and Outlook}
\label{sec:conclusions}

In this article, we presented the package \textsf{SymBuild} which we hope will prove helpful to the community working on the perturbative amplitude bootstrap in computing and manipulating integrable symbols. The package is powerful enough that, even without resorting to \textsf{SpaSM}, the weight four integrable symbols of the non-planar 5-particle alphabet (with 31 letters and one root) can be computed on a laptop in about fifteen minutes.

Regarding future research, several important questions remain open. First, it would be very interesting to understand the geometry and counting behind the integrable symbols more thoroughly. 
For instance, the questions of whether it is possible to devise a formula that would compute the number of integrable symbols of arbitrarily large weight $L$ for a given arbitrary alphabet (even one that is purely rational) is, to our knowledge, unanswered outside of the cases studied in \cite{Brown:2009qja}. 

Somewhat related to the question of counting the number of integrable symbols is the problem of determining the theoretical limits of the applicability of the amplitude bootstrap method itself. Since the number of symbols grows rapidly with the weight, so too does the number of coefficients that need to be determined via consistency with collinear limits, soft limits and others. Disregarding practical concerns on computing time and memory, is it possible to prove that the bootstrap program with succeed at any loop (and hence weight) order or can one determine a priori at which loop order the method will fail?

More practically, it would be interesting to extend this package and the methods therein to the case of elliptic functions \cite{Broedel:2018iwv} or even to the higher generalizations discussed in \cite{Bourjaily:2017bsb}. Furthermore, the questions of which roots can arise in practical cases (and how to resolve them if possible) and of the a priori determination of the alphabet for a given scattering process are still quite open. We refer to \cite{Prlina:2018ukf} for work on the latter problem.

\section*{Acknowledgments}
We are indebted to Marco Besier, Robin Br\"user, Huanhang Chi, Dmitry Chicherin, Lance Dixon, Alessandro Georgoudis, Johannes Henn, Enrico Herrmann, Kasper Larsen, Tiziano Peraro, Pascal Wasser and Simone Zoia for numerous useful discussion, helpful suggestions and assistance with various \textsf{Mathematica} features. We are particularly thankful to Simone Zoia for pointing out numerous bugs as well as improvement possibilities.  The authors are supported in part by the PRISMA Cluster of Excellence at Mainz university. The research leading to these results has also received funding from Swiss National Science Foundation (Ambizione grant PZ00P2 161341), and from the European Research Council (ERC) under the European Union’s Horizon 2020 research and innovation programme (grant agreement No 725110). The authors gratefully acknowledge the computing time
granted on the supercomputer Mogon at Johannes Gutenberg University Mainz (hpc.uni-mainz.de).


\appendix



\bibliography{./auxi/biblio}

\providecommand{\href}[2]{#2}\begingroup\raggedright\begin{thebibliography}{10}

\bibitem{Eden:1966dnq}
R.~J. Eden, P.~V. Landshoff, D.~I. Olive, and J.~C. Polkinghorne, {\em {The
  analytic S-matrix}}.
\newblock Cambridge Univ. Press, Cambridge, 1966.

\bibitem{Bern:1994zx}
Z.~Bern, L.~J. Dixon, D.~C. Dunbar, and D.~A. Kosower, {\it {One loop n point
  gauge theory amplitudes, unitarity and collinear limits}},  {\em Nucl. Phys.}
  {\bf B425} (1994) 217--260, [\href{http://arxiv.org/abs/hep-ph/9403226}{{\tt
  hep-ph/9403226}}].

\bibitem{Zagier1991}
D.~Zagier, {\em Polylogarithms, Dedekind Zeta Functions, and the Algebraic
  K-Theory of Fields (part of the book Arithmetic Algebraic Geometry)},
  pp.~391--430.
\newblock Birkh{\"a}user Boston, Boston, MA, 1991.

\bibitem{Goncharov_1995}
A.~Goncharov, {\it Geometry of configurations, polylogarithms, and motivic
  cohomology},  {\em Advances in Mathematics} {\bf 114} (sep, 1995) 197--318.

\bibitem{Lewin1991}
L.~Lewin, ed., {\em Structural Properties of Polylogarithms}.
\newblock American Mathematical Society, October, 1991.

\bibitem{Gangl_2003}
H.~Gangl, {\it Functional equations for higher logarithms},  {\em Selecta
  Mathematica, New Series} {\bf 9} (sep, 2003) 361--377.

\bibitem{Goncharov:2010jf}
A.~B. Goncharov, M.~Spradlin, C.~Vergu, and A.~Volovich, {\it {Classical
  Polylogarithms for Amplitudes and Wilson Loops}},  {\em Phys. Rev. Lett.}
  {\bf 105} (2010) 151605, [\href{http://arxiv.org/abs/1006.5703}{{\tt
  arXiv:1006.5703}}].

\bibitem{Dixon:2011nj}
L.~J. Dixon, J.~M. Drummond, and J.~M. Henn, {\it {Analytic result for the
  two-loop six-point NMHV amplitude in N=4 super Yang-Mills theory}},  {\em
  JHEP} {\bf 01} (2012) 024, [\href{http://arxiv.org/abs/1111.1704}{{\tt
  arXiv:1111.1704}}].

\bibitem{Dixon:2011pw}
L.~J. Dixon, J.~M. Drummond, and J.~M. Henn, {\it {Bootstrapping the three-loop
  hexagon}},  {\em JHEP} {\bf 11} (2011) 023,
  [\href{http://arxiv.org/abs/1108.4461}{{\tt arXiv:1108.4461}}].

\bibitem{Dixon:2013eka}
L.~J. Dixon, J.~M. Drummond, M.~von Hippel, and J.~Pennington, {\it {Hexagon
  functions and the three-loop remainder function}},  {\em JHEP} {\bf 12}
  (2013) 049, [\href{http://arxiv.org/abs/1308.2276}{{\tt arXiv:1308.2276}}].

\bibitem{Dixon:2014iba}
L.~J. Dixon and M.~von Hippel, {\it {Bootstrapping an NMHV amplitude through
  three loops}},  {\em JHEP} {\bf 10} (2014) 065,
  [\href{http://arxiv.org/abs/1408.1505}{{\tt arXiv:1408.1505}}].

\bibitem{Dixon:2015iva}
L.~J. Dixon, M.~von Hippel, and A.~J. McLeod, {\it {The four-loop six-gluon
  NMHV ratio function}},  {\em JHEP} {\bf 01} (2016) 053,
  [\href{http://arxiv.org/abs/1509.08127}{{\tt arXiv:1509.08127}}].

\bibitem{Caron-Huot:2016owq}
S.~Caron-Huot, L.~J. Dixon, A.~McLeod, and M.~von Hippel, {\it {Bootstrapping a
  Five-Loop Amplitude Using Steinmann Relations}},  {\em Phys. Rev. Lett.} {\bf
  117} (2016), no.~24 241601, [\href{http://arxiv.org/abs/1609.00669}{{\tt
  arXiv:1609.00669}}].

\bibitem{Drummond:2014ffa}
J.~M. Drummond, G.~Papathanasiou, and M.~Spradlin, {\it {A Symbol of
  Uniqueness: The Cluster Bootstrap for the 3-Loop MHV Heptagon}},  {\em JHEP}
  {\bf 03} (2015) 072, [\href{http://arxiv.org/abs/1412.3763}{{\tt
  arXiv:1412.3763}}].

\bibitem{Dixon:2016nkn}
L.~J. Dixon, J.~Drummond, T.~Harrington, A.~J. McLeod, G.~Papathanasiou, and
  M.~Spradlin, {\it {Heptagons from the Steinmann Cluster Bootstrap}},  {\em
  JHEP} {\bf 02} (2017) 137, [\href{http://arxiv.org/abs/1612.08976}{{\tt
  arXiv:1612.08976}}].

\bibitem{Chicherin:2017dob}
D.~Chicherin, J.~Henn, and V.~Mitev, {\it {Bootstrapping pentagon functions}},
  {\em JHEP} {\bf 05} (2018) 164, [\href{http://arxiv.org/abs/1712.09610}{{\tt
  arXiv:1712.09610}}].

\bibitem{Henn:2018cdp}
J.~Henn, E.~Herrmann, and J.~Parra-Martinez, {\it {Bootstrapping two-loop
  Feynman integrals for planar $N=4$ sYM}},
  \href{http://arxiv.org/abs/1806.06072}{{\tt arXiv:1806.06072}}.

\bibitem{Chicherin:2018wes}
D.~Chicherin, J.~M. Henn, and E.~Sokatchev, {\it {Implications of nonplanar
  dual conformal symmetry}},  \href{http://arxiv.org/abs/1807.06321}{{\tt
  arXiv:1807.06321}}.

\bibitem{Almelid:2017qju}
{\O}.~Almelid, C.~Duhr, E.~Gardi, A.~McLeod, and C.~D. White, {\it
  {Bootstrapping the QCD soft anomalous dimension}},  {\em JHEP} {\bf 09}
  (2017) 073, [\href{http://arxiv.org/abs/1706.10162}{{\tt arXiv:1706.10162}}].

\bibitem{Duhr:2012fh}
C.~Duhr, {\it {Hopf algebras, coproducts and symbols: an application to Higgs
  boson amplitudes}},  {\em JHEP} {\bf 08} (2012) 043,
  [\href{http://arxiv.org/abs/1203.0454}{{\tt arXiv:1203.0454}}].

\bibitem{Brown:2009qja}
F.~C.~S. Brown, {\it {Multiple zeta values and periods of moduli spaces M 0 ,n
  ( R )}},  {\em Annales Sci. Ecole Norm. Sup.} {\bf 42} (2009) 371,
  [\href{http://arxiv.org/abs/math/0606419}{{\tt math/0606419}}].

\bibitem{spasm}
The\_SpaSM\_group, {\em {SpaSM}: a Sparse direct Solver Modulo $p$}, v1.2~ed.,
  2017.
\newblock \url{http://github.com/cbouilla/spasm}.

\bibitem{Duhr:2011zq}
C.~Duhr, H.~Gangl, and J.~R. Rhodes, {\it {From polygons and symbols to
  polylogarithmic functions}},  {\em JHEP} {\bf 10} (2012) 075,
  [\href{http://arxiv.org/abs/1110.0458}{{\tt arXiv:1110.0458}}].

\bibitem{Golden:2014xqa}
J.~Golden, M.~F. Paulos, M.~Spradlin, and A.~Volovich, {\it {Cluster
  Polylogarithms for Scattering Amplitudes}},  {\em J. Phys.} {\bf A47} (2014),
  no.~47 474005, [\href{http://arxiv.org/abs/1401.6446}{{\tt
  arXiv:1401.6446}}].

\bibitem{Broedel:2018iwv}
J.~Broedel, C.~Duhr, F.~Dulat, B.~Penante, and L.~Tancredi, {\it {Elliptic
  symbol calculus: from elliptic polylogarithms to iterated integrals of
  Eisenstein series}},  \href{http://arxiv.org/abs/1803.10256}{{\tt
  arXiv:1803.10256}}.

\bibitem{Bourjaily:2017bsb}
J.~L. Bourjaily, A.~J. McLeod, M.~Spradlin, M.~von Hippel, and M.~Wilhelm, {\it
  {Elliptic Double-Box Integrals: Massless Scattering Amplitudes beyond
  Polylogarithms}},  {\em Phys. Rev. Lett.} {\bf 120} (2018), no.~12 121603,
  [\href{http://arxiv.org/abs/1712.02785}{{\tt arXiv:1712.02785}}].

\bibitem{Prlina:2018ukf}
I.~Prlina, M.~Spradlin, and S.~Stanojevic, {\it {All-loop singularities of
  scattering amplitudes in massless planar theories}},
  \href{http://arxiv.org/abs/1805.11617}{{\tt arXiv:1805.11617}}.

\end{thebibliography}\endgroup
\bibliographystyle{./auxi/JHEP}

\end{document}